\documentclass[reprint,twocolumn,longbibliography,eqsecnum,floats,aps,amsmath,amssymb,nofootinbib,notitlepage]{revtex4-2}

\usepackage{graphicx} 
\usepackage{caption}
\usepackage{braket}
\usepackage{subcaption}
\usepackage{tcolorbox}
\usepackage{amsfonts}
\usepackage[mathscr]{eucal}
\usepackage{enumerate}

\usepackage{tikz}
\usepackage{tikz-cd}
\usepackage{tikz}
\usetikzlibrary{shapes,arrows}

\usepackage{amsmath,dsfont}

\newcommand{\rd}{{\rm d}}
\newcommand{\ub}[1]{\underbar{#1}}
\newcommand{\id}{\mathcal{I}}

\definecolor{olivegreen}{RGB}{60,128,49}

\begin{document} 

\title{Revisiting light propagation over (loop) quantum Cosmos}

\author{Aliasghar Parvizi}
\email{a.parvizi@ipm.ir}
\affiliation{Faculty of Mathematics and Computer Science,\\
Transilvania University, Iuliu Maniu Str. 50, 500091 Brasov, Romania}

\author{Tomasz Pawłowski}
\email{tomasz.pawlowski@uwr.edu.pl}
\affiliation{University of Wrocław, Faculty of Physics and Astronomy, Institute of Theoretical Physics, Maksa Borna 9, PL-50-204 Wrocław, Poland}

\begin{abstract}
We investigate the propagation of electromagnetic waves over a quantum cosmological background, aiming to uncover potential signatures of quantum gravity through modifications to the dynamics of the field. Building on symmetry-reduced approaches to spacetime quantization, specifically loop quantum cosmology and geometrodynamics, and extending the Born-Oppenheimer approximation for interacting fields, we construct a quasi-phenomenological framework capable of probing all energy regimes. Unlike previous semi-classical treatments confined to low-energy limits, our analysis employs both analytical and numerical methods to study wave dynamics in a flat quantum Friedmann–Lemaître–Robertson–Walker Universe. Our results confirm consistency with general relativity at low energies, reveal quantum geometric corrections at higher energies, and demonstrate that loop quantum effects suppress modifications relative to those predicted by geometrodynamics-based quantization.
\end{abstract}

\maketitle 

\section{Introduction}

The problem of propagation of quantum fields over a quantum geometry has been attracting considerable attention of researchers, in particular in the context of Loop Quantum Gravity. There, extensive studies have been performed, drawing insights from various approaches and theoretical frameworks \cite{Gambini:1998it, Alfaro:1999wd, Alfaro:2001rb, Ashtekar:2009mb, Agullo:2012sh, Dapor:2013pka}. One notable approach, initially introduced in the work by Ashtekar et al \cite{Ashtekar:2009mb}, involves the concept of a “dressed metric”. In this framework, the inhomogeneous matter field is decomposed onto a set of Fourier modes, each being then a global degree of freedom living on a cosmological (isotropic) background. Then, the classical Friedmann-Lemaître-Robertson-Walker (FLRW) spacetime emerges as an effective metric probed by the quantum test field modes. However, in the case of a massless scalar test field, it has been demonstrated that all modes experience a unified dressed metric. Subsequent studies on the behavior of massive modes, \cite{Assanioussi:2016yxx} found each mode ``feeling'' a distinct classical space-time, leading to the emergence of what is often referred to as a “rainbow metric” \cite{Magueijo:2002xx, Lafrance:1994in}.
Later, this framework has been used to incorporate a form of Born-Oppenheimer approximation motivated by the description of the interaction of light and matter in condensed matter physics in order to probe the backreaction of matter field. Studies have revealed that the components of the emergent dressed metric become intricately linked to the energy of the field modes \cite{Lewandowski:2017cvz, Parvizi:2021ekr}. This connection ultimately gives rise to what we refer to as a ``dressed rainbow metric'', analogous to the chromatic dispersion observed in materials. 

In the present work, our focus lies solely on massless fields, which induce subtle alterations in the fabric of spacetime. We treat these fields as mere perturbations affecting the wavefunction and energy of the system. Importantly, our interest lies in understanding these minor variations in the wavefunction and energy, rather than the ultimate state of the system itself. Consequently, we refrain from attempting a complete solution to the evolution equation, opting instead for a perturbative approach and the application of time-independent response theory.
Rather than seeking direct corrections from quantum gravity models, our approach involves investigating their effects through back-reactions. By doing so, we incorporate amplification parameters into the correction terms. This concept is analogous to the wavelength dependence of linear susceptibility observed in solids, which is directly proportional to the number of atoms in the solid and the modes of the electromagnetic field \cite{Boyd:2008eba}.
Similarly, in the context of light-geometry interactions, we can draw a parallel to the chromatic dispersion effect of light-matter system. Although quantum gravity effects are challenging to observe directly, there exists the possibility of indirect detection through some mechanism that amplifies extremely small Planckian corrections \cite{Amelino-Camelia:2008aez}. 

The concept of modified dispersion relations, which emerge as a consequence of quantum gravity effects, can be traced back to the seminal work of Amelino-Camelia et al. \cite{Amelino-Camelia:1996bln}. In their pioneering study, a semi-classical Liouville string was utilized as a model to explore quantum gravity. Subsequent research has expanded in this direction, in particular, through exploration of quantum deformations of the Poincaré symmetry group via various approaches \cite{Magueijo:2001cr, Kowalski-Glikman:2001vvk, Kowalski-Glikman:2004fsz,Amelino-Camelia:2000cpa,Amelino-Camelia:2000stu,Magueijo:2002am}. The phenomenological aspects of these proposals have been examined in \cite{Amelino-Camelia:1997ieq}.
Following the initial research, a succession of studies has been undertaken to establish a theoretical basis for the manifestation of such effects within the realm of quantum gravity theories. For example, Gambini et al. \cite{Gambini:1998it} explored a theoretical model within which they deduced a modified set of nonlinear Maxwell’s equations, attributing these alterations to the granular nature of spacetime as posited by Loop Quantum Gravity (LQG) within a semi-classical context. This model was further refined in \cite{Alfaro:2001rb}, which continued to work within the semi-classical domain to derive modified Maxwell equations tailored for flat spacetimes.
Progressing along this line, Magueijo et al.  \cite{Magueijo:2002xx} introduced a generalization of Nonlinear Special Relativity, dubbed ``Gravity's Rainbow'', designed to include the concept of curvature. This model maintained its status as an effective theory. In a more recent advancement, \cite{Assanioussi:2014xmz} embarked on the foundational principles of Loop Quantum Cosmology (LQC) to formulate a mode-dependent dressed spacetime, applicable exclusively to massive fields. This was achieved by comparing Quantum Field Theory (QFT) on classical backgrounds with QFT on quantum FLRW geometries in a non-trivial way.

The particular approach employed here originated with the development of the framework in \cite{Lewandowski:2017cvz}, was subsequently refined and extended in \cite{Parvizi:2021ekr}, and is further reexamined in this paper, where we comprehensively revisit previous results concerning light propagation. In addition, we explicitly calculate the modifications of the dispersion relation across energy scales within two distinct approaches to quantum gravity, verifying their mutual consistency. We also perform a stability analysis, establish convergence criteria, and conduct an operator variance analysis for the numerical computation. The strength of this approach lies in its relative robustness and its well-defined systematic structure.
Contrary to previous studies which were primarily grounded in ad hoc assumptions and remained largely analytical within the semi-classical regime, our approach originates from a fundamental theory. We introduce a robust framework for examining light--geometry interactions and carry out comprehensive analyses and consistency checks across distinct regimes: from the deep quantum regime, through the semi-classical regime, and finally to the classical regime, employing both analytical and numerical methods.

Before presenting the details of our analysis, we briefly recall the basic mechanisms of light–matter (L--M) interaction and the origin of chromatic aberration in materials. A more extensive review and comparison is provided in Appendix~\ref{app-CDMMaterials}.
To bridge the fundamental ideas underlying light–matter interactions with those relevant for light–geometry (L--G) interactions, and to justify our treatment of backreaction and the emergence of chromatic dispersion in quantum geometry, we draw explicit parallels with analogous phenomena in material science.
In conventional media, the derivation of optical susceptibility relies on the interaction between atoms and the electromagnetic field. This is typically formulated using quantum‑mechanical perturbation theory applied to the atomic wave function, together with linear response theory \cite{solyom2008fundamentals, Boyd:2008eba}. In this context, the total Hamiltonian for the light-matter system is
\begin{equation}
    \widehat{H} = \widehat{H}_0 + \widehat{V} + \widehat{H}_{EM}
\end{equation}
where $\widehat{H}_0$ denotes the Hamiltonian of a free atom and $\widehat{V}$ represents the operator describing the interaction between the atom and the electromagnetic field. The last term, $\widehat{H}_{\rm EM}$, corresponds to the Hamiltonian of the electromagnetic field itself.
When the perturbed quantum state of the atom is constructed, one can compute the electric dipole moment $\langle \mathbf{\widehat{p}} \rangle$ and extract the frequency-dependent linear susceptibility $\chi(\omega_p)$; see Appendix~\ref{app-CDMMaterials} for explicit relations, while further details are contained in \cite{solyom2008fundamentals,Boyd:2008eba}.

The above procedure will be incorporated into our LQC framework and employed to derive the modified dynamics and compute the geometrical observables. A comparison between our approach in the context of a L--M system and that of a L--G system reveals both notable similarities and crucial differences. In both cases, we work with first-order perturbations and solve the corresponding equations to obtain $\psi_o(v, T)$ (unperturbed state) and $\tilde{\psi}(v, T)$ (perturbed state) in the L--G system. Similarly to the L--M system, we mode-expand the interaction Hamiltonian as $\widehat{H}_{EM} = \Sigma_{r} \Sigma_{\mathbf{k}} \widehat{H}_{\mathbf{k}}^{(r)}$. Within the linear response regime, we calculate the geometric quantities $\langle\widehat{V}^n\rangle$ with respect to the geometry quantum state $\tilde{\psi}(v, T)$, analogous to the evaluation of $\langle \mathbf{\widehat{p}} \rangle = \langle \tilde{\psi} | \mathbf{\widehat{\mu}} | \tilde{\psi} \rangle$ in the L-M system. Although our approach shares similarities with the L--M system, there are also key differences. In the L--M system, the solutions for the electromagnetic Hamiltonian are known, whereas in the L--G system, they must be determined based on the state of the geometry. This introduces an additional layer of complexity into our analysis. Furthermore, in the L--M system, an extra term representing the interaction between light and matter must be added to the atomic Hamiltonian. In contrast, in the L--G system, the electromagnetic Hamiltonian $\widehat{H}_{EM}$ itself serves as the interaction term, composed of both geometric and matter operators. 

The article is organized as follows. We begin by introducing the investigated model and outlining the employed method, namely a modification of the Born--Oppenheimer approximation, in Sec.~\ref{quantum-sys}. In 
Sec.~\ref{app-numerics}, we perform a numerical analysis of the system and 
construct the perturbed quantum geometry state $\tilde{\psi}(v,T)$. 
Sec.~\ref{sec:eff} presents and describes the studied system within the 
so-called zeroth-order effective dynamics formalism. The content of these 
sections is then used to establish a systematic procedure for obtaining the 
dispersion relation governing the propagation of electromagnetic field modes, as detailed in Sec.~\ref{sec:dispersion}. In Sec.~\ref{sec:props-DR}, both analytical and numerical constructions are applied to determine the geometrical observables and study their properties for a population of states representing wave propagation in a quantum universe. The results and their implications are then discussed in Sec.~\ref{sec:results}, together with the concluding remarks.

\section{Quantization and Born-Oppenheimer approximation}
\label{quantum-sys}

We begin by specifying the system under study. 
In our analysis, we consider a system described by a metric representing a flat (isotropic) Friedmann-Lemaitre-Robertson-Walker (FLRW) universe, $ds^2 = - N^2(t) dt^2 + a^2(t) d\mathbf{x}^2$, coupled with irrotational dust providing a material time reference, as discussed in ~\cite{Husain:2011tk}. In that system, we incorporate $U(1)$ symmetry perturbations (an e-m field) on the background metric, as was also explored in Ref.~\cite{Lewandowski:2017cvz}.
The corresponding action for the system reads:
\begin{equation}
S\ =\  \int d^4x  \left[\frac{\sqrt{-g}}{8\pi G}\mathcal{R} + \mathcal{L}_{\rm T}\right] + S_{\rm EM}\ .
\label{action-tot}
\end{equation}
Here, $\mathcal{R}$, $\mathcal{L}_{\rm T}$, and $S_{\rm EM}$ denote the 
gravity Lagrangian density, the dust Lagrangian density, and the action of the electromagnetic field, respectively. Then the total Hamiltonian will be given by
\begin{equation}
H\ =\ \int d^3x \big[\mathcal{H}_{\rm gr} + \mathcal{H}_{\rm T} +\mathcal{H}_{\rm EM}\big],
\label{Ham-tot}
\end{equation}
where $\mathcal{H}_{\rm gr}$, $\mathcal{H}_{\rm T}$, and $\mathcal{H}_{\rm EM}$ are,  respectively, the Hamiltonian density of the gravitational, dust, and e-m field sectors. Since in the system under consideration the total Hamiltonian satisfies \(\mathcal{H}_{\rm T}=p_T\), where \(p_T\) is the momentum conjugate to the dust field \(T\) and obeys the canonical relation \(\{T,p_T\}=1\) \cite{Husain:2011tk}, the dust field can be used as the evolution parameter. This allows us to deparametrize the theory and treat the evolution with respect to \(T\) in a consistent way. This leads to
\begin{equation}\label{trueH}
    \begin{split}
      -p_T &=: \mathbb{H} \, , \\
     \mathbb{H} &\equiv H_{\rm gr} + H_{\rm EM} \, ,        
    \end{split}
\end{equation}
in which we introduce the true Hamiltonian $\mathbb{H}$ with respect to the new time variable $T$. Using the Fourier expansion, we can rewrite $H_{\rm EM}$ as an assembly of the Hamiltonians of decoupled harmonic oscillators $H_{\rm EM} = \sum_{\mathbf{k}\in{\cal L}} \sum_{r}^2 H_{\mathbf{k}}^{(r)}$, each represented by a pair of canonically conjugate variables $(Q_{\mathbf{k}}^{(r)}, P_{\mathbf{k}}^{(r)})$ \cite{Lewandowski:2017cvz}
\begin{equation}\label{Hamiltonian-SF1}
H_{\mathbf{k}}^{(r)} =  \frac{N(T)}{2a^3(T)} 
\Big[\big(P_{\mathbf{k}}^{(r)}\big)^2 + k^2a^4(T)\big(Q_{\mathbf{k}}^{(r)}\big)^2\Big] \ ,
\end{equation}
for each mode $\mathbf{k}$ and polarization $(r)$ of the e-m field, where they satisfy the relation
$\{Q_{\mathbf{k}}^{(r)}, P_{\mathbf{k}^\prime}^{(r^\prime)}\}=\delta_{\mathbf{k}\mathbf{k}^\prime}\delta_{rr^\prime}$. The wave vector $\mathbf{k}$ belongs to the lattice
$\mathcal{L} := \left(\tfrac{2\pi}{\ell}\,\mathbb{Z}\right)^3$, where $\mathbb{Z}$ denotes the set of integers. By imposing the canonical time gauge fixing condition, i.e., $N(t)=1$, the gravitational Hamiltonian reads \cite{Parvizi:2021ekr}
\begin{equation}\label{Hgr}
H_{\rm gr} = \int d^3x \mathcal{H}_{\rm gr} =\  \frac{3\pi G}{2\alpha_o}b^2|v| \, .
\end{equation}
The phase space of the model is conveniently described using the canonically conjugate pair \(\{b, v\} = 2\). The variable \(v\) encodes the oriented physical volume through \(|v| = a^{3}/\alpha_{0}\), while \(b = \gamma\,(\dot{a}/a)\) represents a rescaled Hubble parameter, with the overdot denoting differentiation with respect to the dust proper time. The constant \(\gamma\) is the Barbero–Immirzi parameter characteristic of loop quantum gravity, and  $\alpha_{0} = 2\pi\gamma\sqrt{\Delta}\,\ell_{\rm Pl}^{2}$, where \(\Delta = 4\sqrt{3}\,\pi\gamma\,\ell_{\rm Pl}^{2}\) is the fundamental {\em area gap} of loop quantum cosmology and \(\ell_{\rm Pl}\) is the Planck length \cite{Ashtekar:2006rx}. The sign of \(v\) keeps track of the triad orientation, ensuring that the phase‑space description faithfully captures both geometric magnitude and orientation.
This parametrization is particularly advantageous in LQC because it isolates the variables that directly participate in the quantization procedure of operators--namely, the holonomy‑related quantity \(b\) and the discrete volume label \(v\). 

Following the Dirac quantization scheme for the constrained systems, a total kinematical Hilbert space  for the above geometry-matter system  can be defined as $\mathscr{H}_{\rm kin} = \mathscr{H}_{\rm gr} \otimes \mathscr{H}_{T}  \otimes \mathscr{H}_{\rm EM}$. The e-m Hamiltonian on the cosmological background can be broken down into a collection of harmonic oscillators. Each mode corresponds to a Hilbert space $\mathscr{H}^{(r)}_{\mathbf{k}}$ and an associated Hamiltonian operator $\widehat{H}_{\mathbf{k}}^{(r)}$ \cite{Lewandowski:2017cvz}. At the next step, matter sectors are quantized according to the Schr\"odinger picture with the Hilbert spaces $\mathscr{H}_{T}=L^2 (\mathbb{R}, dT)$ and $\mathscr{H}_{\mathbf{k}}^{(r)}=L^2( \mathbb{R}, dQ_{\mathbf{k}}^{(r)})$. The corresponding quantum operators on $\mathscr{H}_{\rm kin}$ are those acting on the physical states $\Psi_{\mathbf{k}}(v, Q_{\mathbf{k}}^{(r)}, T)\in \mathscr{H}_{\rm kin}$. The quantization is carried out in the same way as for the quantum harmonic oscillator with the phase space variables promoted to operators on the corresponding Hilbert spaces, as 
\begin{subequations}\label{def:operators}
    \begin{align}
        \widehat{Q}^{(r)}_\mathbf{k} \, \Psi_{\mathbf{k}}(v, Q^{(r)}_\mathbf{k}, T) &=  {Q}^{(r)}_\mathbf{k} \, \Psi_{\mathbf{k}}(v, Q^{(r)}_\mathbf{k}, T) ,\\
        \widehat{P}^{(r)}_\mathbf{k} \Psi_{\mathbf{k}}(v, Q^{(r)}_\mathbf{k}, T) &= -i\hbar \partial/\partial {Q}^{(r)}_\mathbf{k} \, \Psi_{\mathbf{k}}(v, Q^{(r)}_\mathbf{k}, T),
    \end{align}
\end{subequations}
and 
\begin{subequations}\label{def:operators2}
    \begin{align}
        \widehat{T} \, \Psi_{\mathbf{k}}(v, Q^{(r)}_\mathbf{k}, T) &=  T \, \Psi_{\mathbf{k}}(v, Q^{(r)}_\mathbf{k}, T), \\
        \widehat{p}_T \Psi_{\mathbf{k}}(v, Q^{(r)}_\mathbf{k}, T) &= -i\hbar \partial/\partial T \, \Psi_{\mathbf{k}}(v, Q^{(r)}_\mathbf{k}, T).
    \end{align}
\end{subequations}

To quantize the gravity sector, we use the LQC framework. LQC \cite{Ashtekar:2011ni} is an approach to quantum gravity description of spacetimes relevant for cosmology via the application of the quantization methods of LQG to reduced models of highly symmetric spacetimes. In the process of implementing the constraint following the Dirac program, the gravitational Hamiltonian will be reexpressed in terms of holonomies and fluxes using Thiemann regularization \cite{Thiemann:2007pyv}. However, due to the physical meaning of the embedding, the framework no longer operates within the diffeomorphism-invariant sector. As a result, when reexpressing the curvature through an appropriate limit of holonomies over a closed loop, one can no longer take the limit where the loop shrinks to a point. Instead, loops are fixed by requiring that their physical area corresponds to the lowest nonzero eigenvalue of the full LQG area operator \cite{Ashtekar:2006wn}. This serves as a semi-phenomenological input from the full theory, although arguments exist supporting its validity--at least at the level of magnitude (see, for example, \cite{Pawlowski:2014nfa}). Thus, the geometry degrees of freedom are quantized in the LQC framework (see \cite{Ashtekar:2003hd}, \cite{Ashtekar:2006wn}, \cite{Ashtekar:2011ni} for more details on the framework) with the Hilbert space $\mathscr{H}_{\rm gr}=L^2 (\bar{\mathbb{R}}, d\mu_{\rm Bohr})$, where $\bar{\mathbb{R}}$ is the Bohr compactification of the real line and $d\mu_{\rm Bohr}$ is the Haar measure on it \cite{Ashtekar:2003hd}. In this scheme, the gravitational quantum Hamiltonian \eqref{Hgr} becomes \cite{Husain:2011tm}
\begin{equation} \label{Ham_grav}
\widehat{H}_{\rm gr} = \frac{3\pi G}{8\alpha_o} \sqrt{\hat{v}} \big(\widehat{\cal{N}}^2 - \widehat{\cal{N}}^{-2}\big)^2 \sqrt{\hat{v}} \, ,
\end{equation}
where $\hat{v}|v\rangle=v|v\rangle$ and the operator $\widehat{\cal{N}}\equiv\widehat{\exp(ib/2)}$ act on the  basis $\{|v\rangle\}$, i.e. the eigenstates of $\hat{v}$, as  $\widehat{\cal{N}}|v\rangle = |v +1\rangle$, so that $[\hat{b}, \hat{v}]=2i\hbar$. Upon promoting the relevant fundamental components to operators, the Hamiltonian becomes a difference operator of low order, usually second \cite{Parvizi:2021ekr}. In which $\hat{v}$ is the oriented volume and is related to the physical volume of the universe $V$ by $V = \ell^3 a^3 = \alpha_0 |v|$, where $\ell$ is the coordinate length of the comoving fiducial cell.

After quantizing the Hamiltonian constraint Eq.~\eqref{trueH},
\begin{equation}\label{trueH-Q}
          -\widehat{p}_T = \widehat{H}_{\rm gr} + \widehat{H}_{\mathbf{k}}^{(r)} ,
\end{equation}
for one mode and polarization of e-m field, we obtain a Schr\"odinger-like equation
\begin{equation}
    i\hbar \partial_T \Psi_{\mathbf{k}}(v, Q_{\mathbf{k}}^{(r)}, T) 
    = \Big[\widehat{{H}}_{\rm gr}  + \widehat{H}_{\mathbf{k}}^{(r)}  \Big] \Psi_{\mathbf{k}}(v, Q_{\mathbf{k}}^{(r)}, T), 
\label{QG-Shroedinger-2}
\end{equation}
for the quantum gravity state $\Psi_{\mathbf{k}}(v, Q_{\mathbf{k}}^{(r)}, T)$, where the quantum operator counterpart of the e-m Hamiltonian \eqref{Hamiltonian-SF1} reads as
\begin{align}\label{eq:Hem}
    \widehat{H}_{\mathbf{k}}^{(r)} &= \frac{1}{2} \big[\alpha_0^{-1} |\hat{v}|^{-1} \otimes \left(\widehat{P}^{(r)}_\mathbf{k}\right)^2 \nonumber\\
    &+ \ell^{-4} k^2 \alpha_0^{1/3} |\hat{v}|^{1/3} \otimes \left(\widehat{Q}^{(r)}_{\mathbf{k}}\right)^2\big], 
\end{align}
which is composed of geometric and field operators, in which we used $\widehat{V} = \ell^3 \hat{a}^3 = \alpha_0 |\hat{v}|$ and set $N(T) = 1$.

Before we explore the specifics, let's outline our program to facilitate a clearer understanding for the readers. Similar to the derivation of mode-dependent susceptibility, we operate at the linear order and compute the back-reaction of each mode independently within the Born–Oppenheimer (BO) approximation for the system \eqref{QG-Shroedinger-2}. Various techniques can be employed to account for the  backreaction effect. One such approach involves the BO decomposition applied to the Hilbert space. For an in-depth exploration in the context of quantum gravity, refer to \cite{Kiefer:2004xyv} and \cite{Rovelli:2008aa} . Additionally, the application of this approximation method within the spinfoam approach to LQC is explored in \cite{Rovelli:2008aa} and discussed in the context of the hybrid approach to LQC in \cite{CastelloGomar:2016rjj}. The BO methods for LQG and the necessary conditions that the gravitational field algebra must satisfy are elucidated in \cite{Giesel:2009at}. Additionally, a more extensive generalization of the BO approximation, applicable to a diverse range of quantum systems, is explored in \cite{Schander:2019wuq, Schander:2019ovh, Schander:2019bca} and reviewed in \cite{Schander:2021pgt}. Born-Oppenheimer approximation assumes that the wave functions of geometry and e-m field in our system can be treated separately. This is based on the fact that geometry degrees of freedom are heavier than e-m modes due to their larger relative energy difference. Consequently, we can solve the eigenvalue problem of the entire system perturbatively and compute the backreaction of e-m modes on the background geometry quantum state.
To employ the BO approximation, we begin by making the ansatz: $\Psi_\mathbf{k}(v, Q_\mathbf{k}^{(r)}, T) = \psi_\mathbf{k}(v, T) \otimes \phi_\mathbf{k}(Q_\mathbf{k}^{(r)}, T;v)$ for the total wave function of the light-geometry system, where $\psi(v, T)$ and $\phi_\mathbf{k}(Q_\mathbf{k}^{(r)}, T;v)$ are the wave functions of geometry and e-m modes, respectively. 
In contrast to the scenario encountered in the interaction between e-m fields and atoms within materials, the e-m Hamiltonian, denoted as \eqref{Hamiltonian-SF1}, fundamentally represents an interacting Hamiltonian, which is intricately entwined with geometrical operators.
For the Born–Oppenheimer approximation to work, the dependence of $\phi_\mathbf{k}(Q_\mathbf{k}^{(r)}, T;v)$ on $v$ should be interpreted in a parametric manner, specifically, we should hold $v$ constant and treat $\phi_\mathbf{k}(Q_\mathbf{k}^{(r)}, T;v)$ as an element within the Hilbert space $\mathscr{H}_{\rm EM}$.
This assumption is justified when working in a representation where $\hat{v}$ acts as a multiplication operator; furthermore, the energy scales of the matter and gravity parts are well separated \cite{Giesel:2009at}. We now start by solving the eigenvalue equation
\begin{equation}\label{eq:eigenfield}
     \widehat{H}_{\mathbf{k}}^{(r)} \chi^n_\mathbf{k} (Q^{(r)}_\mathbf{k};v) = e^n_\mathbf{k}({v}) \, \chi^n_\mathbf{k}(Q^{(r)}_\mathbf{k};v) ,
\end{equation}
where $\chi^n_\mathbf{k} (Q^{(r)}_\mathbf{k};v) $ are the field eigenvectors that span the field's state $\phi_\mathbf{k} (Q^{(r)}_\mathbf{k}, T;v) $. It can be solved to give \cite{Parvizi:2021ekr}, 
\begin{subequations}\begin{align}
&{\epsilon}^n_{\mathbf{k}}  = \left(n+\frac{1}{2}\right){V}^{-1/3} \, \hbar k \ell^{-2} \, , \label{epsilon-2} \\
&\chi^n_{\mathbf{k}}(Q^{(r)}_\mathbf{k};v) =     
a_n \left(\frac{k}{\hbar}\right)^{\frac{1}{4}} {V}^{1/12}
 \exp \left(-\frac{k{V}^{1/3}}{2\hbar}(Q_{\mathbf{k}}^{(r)})^2\right)\, \nonumber\\
 &\times H_n\left(\sqrt{\frac{k}{\hbar}} {V}^{1/6} Q_{\mathbf{k}}^{(r)}\right).  
\label{field-sol2}
\end{align}\end{subequations}
In the next step, we return to equation \eqref{QG-Shroedinger-2} and left-multiply it by $\phi_\mathbf{k}(Q_\mathbf{k}^{(r)}, T;v)$ and take the scalar product in the space $\mathscr{H}_{\rm EM}$, which gives the corresponding geometry eigenvalue equation,
\begin{equation}\label{eq:H-corr}
     [\widehat{H}_{\rm gr} + e^n_\mathbf{k}(\hat{v})] \xi^\mu_\mathbf{k}(v) = E^\mu_\mathbf{k} \xi^\mu_\mathbf{k}(v) \ , 
\end{equation}
To obtain the above equation, we disregarded all off-diagonal terms in the matrix elements of  $\widehat{H}_{\rm gr}$ resulting from the action of the geometry operator $\widehat{\cal{N}}$ on $\phi_\mathbf{k}(Q_\mathbf{k}^{(r)}, T;v)$. This simplification is valid when $| e^n_\mathbf{k}({v}) - e^m_\mathbf{k}({v}) | \gg |E^\mu_\mathbf{k} - E^\nu_\mathbf{k}|$.  Essentially, this means that the impact of light degrees of freedom on the dynamics of geometric heavy degrees of freedom is effectively captured solely by their eigenvalues, where $e^n_\mathbf{k}(\hat{v})$ encodes the energy of a single mode of the (test) field state and equals
\begin{equation}\label{e-m-eigenE}
     e^n_\mathbf{k}(\hat{v}) = (n+1/2)\hbar \, k\, \ell^{-2} \, \widehat{V}^{-1/3}  \ . 
\end{equation}
The spectrum of the eigenvalue problem \eqref{eq:H-corr} is nondegenerate and continuous and consists of the entire real line. The eigenvectors $\xi^\mu_\mathbf{k}(v)$ can be found by numerical means via methods used in Refs.~\cite{Ashtekar:2006wn,MenaMarugan:2011me}.
The properties of $\xi^\mu_\mathbf{k}$ and of the 1st order corrected Hamiltonian in \eqref{eq:H-corr} have been investigated in detail in \cite{Parvizi:2021ekr}. In particular, the (essential part of the) spectrum of this Hamiltonian equals $\mathbb{R}^+$, and the eigenfunctions themselves retain the crucial properties of their background geometry counterparts (reflected wave behavior). In particular, their large $v$ asymptotics have been investigated in detail in Appendix B of \cite{Parvizi:2021ekr}. This asymptotic behavior provides us with a precise definition of the label $\mu$ as the energy of dust field.
The form of the asymptotic also implies that $\xi^\mu_{\mathbf{k}}$ are Dirac-delta normalizable. We can thus form out of them an orthonormal basis (for each value of ${\mathbf{k}}$ independently), setting
\begin{equation}
(\xi^\mu_{\mathbf{k}}|\xi^{\mu'}_{\mathbf{k}}) \simeq \delta(\mu-\mu'),
\end{equation}
which allows one to directly apply the numerical methods of LQC (see for example \cite{MenaMarugan:2011me, Ashtekar:2006rx}) to construct solutions to \eqref{QG-Shroedinger-2} corresponding to a chosen spectral profile $c_{\mathbf{k}}(\mu)$ \cite{Parvizi:2021ekr} 
\begin{equation}
\tilde{\psi}_\mathbf{k}(v, T) =  \int_{\mu \in \mathbb{R}} d\mu\,  c_{\mathbf{k}}(\mu)\,  \xi^\mu_{\mathbf{k}}(v) \, e^{iE^{\mu}_{\mathbf{k}} T}. \qquad 
\label{BO1}
\end{equation}
If we trace out the geometrical degrees of freedom in Eq.~\eqref{QG-Shroedinger-2}, using the unperturbed geometric state $\psi_o(v, T)$, we find the Schrodinger-like equation
\begin{equation}\begin{split}
    &i\hbar\partial_T\phi^{(r)}_\mathbf{k}(Q^{(r)}_\mathbf{k}, T) \\ 
    &= \frac{1}{2}\Big[\langle\hat{V}^{-1}\rangle_o\, \big(\hat{P}^{(r)}_\mathbf{k}\big)^2 + \ell^{-4}k^2\, \langle\hat{V}^{\frac{1}{3}}\rangle_o\, \big(\hat{Q}^{(r)}_\mathbf{k}\big)^2\Big]\phi^{(r)}_{\mathbf{k}}(Q^{(r)}_\mathbf{k}, T),
\end{split}\label{Schroedinger2}
\end{equation}
for the e-m quantum state $\phi^{(r)}_\mathbf{k}(Q^{(r)}_\mathbf{k}, T)$. On the other hand, having found the eigenfunctions $|\xi^\mu_{\mathbf{k}})$ of the perturbed geometry and constructed the perturbed state $\tilde{\psi}_\mathbf{k}(v, T)$, alternatively, we could use this state to trace out geometry degrees of freedom and obtain the following Schrodinger-like equation for each mode and polarization of the e-m field:
\begin{equation}\begin{split}
    &i\hbar\partial_T\phi^{(r)}_\mathbf{k}(Q^{(r)}_\mathbf{k}, T) \\ 
    &= \frac{1}{2}\Big[\langle\hat{V}^{-1}\rangle_\mathbf{k}\, \big(\hat{P}^{(r)}_\mathbf{k}\big)^2 + \ell^{-4}k^2\, \langle\hat{V}^{\frac{1}{3}}\rangle_\mathbf{k}\, \big(\hat{Q}^{(r)}_\mathbf{k}\big)^2\Big]\phi^{(r)}_{\mathbf{k}}(Q^{(r)}_\mathbf{k}, T),
\end{split}
\label{Schroedinger3}
\end{equation}
in which we have defined the expectation values $\langle \cdot \rangle_\mathbf{k}$ with respect to the perturbed state $\tilde{\psi}_\mathbf{k}(v, T)$.
The effective equations (\ref{Schroedinger2}) and (\ref{Schroedinger3}) correspond to the evolution equation of the e-m quantum state $\phi^{(r)}_\mathbf{k}(Q^{(r)}_\mathbf{k}, T)$ on a dressed background \cite{Ashtekar:2009mb,Parvizi:2021ekr}. By comparison to the quantum field theory on a classical spacetime with Hamiltonian \eqref{Hamiltonian-SF1}, we can assign the emerging dressed metrics $d\bar{s}^2 = \bar{g}_{ab} dx^adx^b = - \bar{N}^2(T) dt^2 + \bar{a}^2(T) d{\bf x}^2$ and $d\tilde{s}^2 = \tilde{g}_{ab} dx^a dx^b = - \tilde{N}^2(T) dt^2 + \tilde{a}^2(T) d{\bf x}^2$ to the evolution equations (\ref{Schroedinger2}) and (\ref{Schroedinger3}), which are composed of the expectation values of geometric quantum operators. Thus, we find the following relations between the components of the dressed metric and the expectation values of operators with respect to the quantum state ${\Psi}_o(v,T)$:
\begin{equation}\begin{split}
\bar{N}_T(T) &=  \ell^{-3} \left[\big\langle\hat{V}^{-1}\big\rangle_o ~\big\langle\hat{V}^{1/3}\big\rangle_o^3 \right]^{\frac{1}{4}}, \\
\bar{a}(T) &= \left[\big\langle\hat{V}^{1/3}\big\rangle_o\, \big\langle\hat{V}^{-1}\big\rangle_o^{-1}\right]^{\frac{1}{4}},\label{dressed-metric-eq2a}
\end{split}\end{equation}
and $\tilde{\psi}_\mathbf{k}(v,T)$ 
\begin{equation}\begin{split}
\tilde{N}_T(T) &=  \ell^{-3} \left[\big\langle\hat{V}^{-1}\big\rangle_\mathbf{k} ~\big\langle\hat{V}^{1/3}\big\rangle_\mathbf{k}^3 \right]^{\frac{1}{4}}, \\
\tilde{a}(T) &= \left[\big\langle\hat{V}^{1/3}\big\rangle_\mathbf{k}\, \big\langle\hat{V}^{-1}\big\rangle_\mathbf{k}^{-1}\right]^{\frac{1}{4}},
\label{dressed-metric-eq2-BO2}
\end{split}\end{equation}
for the unperturbed and perturbed backgrounds, respectively. 
Figure~\ref{fig:schem} presents a schematic diagram of the procedure we employ to construct a dressed metric, which serves as the background for the propagation of the electromagnetic field.
\tikzstyle{block} = [rectangle, draw, fill=green!20, text width=7.5em, text centered, rounded corners, minimum height=3em]
\tikzstyle{line} = [draw, -latex']
\begin{figure*}[t]
  \centering
  {\scriptsize
  \begin{tikzpicture}[node distance = 2cm, auto]
    \node [block] (consent) {$-\widehat{p}_T = \widehat{H}_{\rm gr} + \widehat{H}_{\mathbf{k}}^{(r)} $};
    \node [block, above of = consent, node distance = 2.5cm, text width=7em] (screening) {$ i\hbar \partial_T \Psi_o(v, T) = \hat{{H}}_{\rm gr}  \Psi_o(v, T),$};
    \node [block, right of = screening, node distance = 4.2cm, text width=8em] (refer) {$\bar{g}: \bar{N}(T),~ \bar{a}(T)$};
    \node [block, right of = refer, node distance = 4cm, text width=8.8em] (refer2) { $ i\hbar \partial_T \psi_\mathbf{k}(Q_\mathbf{k}, T) = \hat{H}_{T, \mathbf{k}}^{(r)}  \psi_\mathbf{k}(Q_\mathbf{k}, T),$};
    \node [block, right of = refer2, node distance = 4.5cm, text width=8.8em] (refer3) { $     [\hat{H}_{\rm gr} + e^n_\mathbf{k}(\hat{v})] \xi^\mu_\mathbf{k}(v) = E^\mu_\mathbf{k} \xi^\mu_\mathbf{k}(v) \ $};
    \node [block, below of = refer3, node distance = 2.5cm] (refer4) { $\tilde{g}: \tilde{N}(T),~ \tilde{a}(T)$};
    
    \path [line] (consent) -- node[anchor=west] {\textrm{(1)}} (screening);
    \path [line] (screening) -- node {\textrm{(2)}} (refer);
    \path [line] (refer) -- node {\textrm{(3)}} (refer2);
    \path [line] (refer2) -- node {\textrm{(4)}} (refer3);
    \path [line] (refer3) -- node[anchor=east] {(5)} (refer4);
  \end{tikzpicture}}
  \caption{A schematic outlining the procedure for solving the Hamiltonian constraint for a light-geometry system: 1) solve the unperturbed geometry state, 2) find the corresponding dressed metric for the field, 3) solve the eigenvalue equation for the field, 4) substitute the energy eigenvalue into the perturbed equation of the background geometry, and 5) solve the perturbed geometry eigenvalue equation to find the corresponding mode-dependent dressed metric for the field.} \label{fig:schem}
\end{figure*}
\section{Numerical analysis of the loop quantum dynamics}
\label{app-numerics}
The core analysis presented here relies on probing the genuine quantum dynamics of the given system with an extensive use of numerical methods. In presented studies, this level of analysis is performed for two systems: (1) the background system of gravity (isotropic degrees of freedom) couple to dust clock field, and (2) the projection onto (again isotropic) geometry degrees of freedom of the full Hilbert space of geometry and electromagnetic radiation modes in Born-Oppenheimer approximation. In both cases, the resulting system is a Hamiltonian system, where isotropic geometry degrees of freedom evolve with respect to the dust field potential.

Here, we provide an overview of the methodology used in these studies. The numerical treatment parallels that of \cite{Parvizi:2021ekr} and involves two main steps: $(i)$ determining the basis of the physical Hilbert space by solving the eigenvalue problem \eqref{eq:H-corr} and properly normalizing the solutions, and $(ii)$ integrating the wave function profiles at a given time $T$ via \eqref{BO1}, followed by evaluating the expectation values and variances of relevant observables based on these profiles. Both these steps use the methodology and numerical code developed and used for earlier LQC projects, see for example \cite{Ashtekar:2006wn} and in particular \cite{MenaMarugan:2011me} for the presentation of the numerical code library used here.

The particularly convenient choice for the physical Hilbert space basis is the ``energy'' basis defined by the spectral decomposition of the Hamiltonian. As for both the background and backreacted system, the treatment is the same, and we will describe the latter one.
Since the inner product induced by loop quantization is $\langle v|v'\rangle = \delta_{v,v'}$, the geometry Hilbert space is nonseparable; however, the form of the Hamiltonian divides it onto a set of separable superselection sectors of states supported on the lattices ${\cal L}_{\epsilon} = \epsilon+4\mathbb{Z}$ (where the superselection sector is labeled by $\epsilon\in [0,4)$). In early studies of these systems (see \cite{Ashtekar:2006wn} and particularly \cite{Bentivegna:2008bg,MenaMarugan:2011me}), generic sectors were considered. However, no significant physical differences were detected in the results across these sectors. Consequently, later studies focused on the sector $\epsilon=0$, which is also the approach taken here. Furthermore, since the model does not include fermions, the triad parity symmetry $v \mapsto -v$ constitutes a large gauge symmetry. As a result, the analysis is further restricted to this symmetric sector.

The properties of $\widehat{H}_{\rm gr}$ have already been investigated in the literature (see for example \cite{Husain:2011tm}). In the sector distinguished above, its spectrum forms a positive real line and is both continuous and nondegenerate. The backreaction term in \eqref{eq:H-corr} is subleading with respect to the diagonal term in $\widehat{H}_{\rm gr}$ and does not change these properties \cite{Parvizi:2021ekr}. As a consequence, the basis of the (relevant sector of the) Hilbert space can be evaluated as a solution to the generalized eigenvalue equation \eqref{eq:H-corr}, 
\begin{equation}\label{eq:eig-mk2}
  \widehat{H}_{\bf k} \xi^n_{\bf k} = E^{\mu}_{\bf k}\xi^{\mu}_{\bf k} \ , \quad E^{\mu}_{\bf k}\geq 0 \ ,
\end{equation}
which is a $2$nd order difference equation (for its specific form, see \cite{Parvizi:2021ekr}). Since the point $v=0$ decouples from the rest of ${\cal L}_{0}$, the entire eigenfunction $\xi^{\mu}_{\bf k}(v)$ can be uniquely determined by iteratively solving \eqref{eq:eig-mk2} with the initial value $\xi^n_{\bf k}(v=4)$. This procedure, however, determines the eigenstate only up to a normalization. Since $\xi^{\mu}_{\bf k}$ in general is not normalizable--given that the spectrum of $\hat{H}_{\bf k}$ is continuous--careful consideration is required to determine its proper normalization.

To determine the norm of a numerically evaluated eigenfunction, one utilizes the fact that, for large $v$, the function $\xi^{\mu}_{\bf k}(v)$ converges to a specific combination of simpler analytically expressible functions \cite{Parvizi:2021ekr}
\begin{equation}\label{eq:xi-conv}
  \xi^{\mu}_{\bf k}(v) = C \left( e^{i\varphi(\mu,{\rm k})} e^{+}_{\mu,{\rm k}}(v) + e^{-i\varphi(\mu,{\rm k})} e^{-}_{\mu,{\rm k}}(v) \right) + O(v^{-9/4}) \ ,  
\end{equation}
where $\varphi(\mu,{bf k})$ is a phase shift and
\begin{equation}
\begin{split}
  e^{\pm}_{\mu,{\bf k}}(v) &= \frac{|v|^{-1/4}}{\sqrt{4\pi}} \left[1+\sum_{i=1}^5 a_n(\mu,{\bf k}) |v^{-n/3}|\right] \\
  &\exp\left[\pm i\mu|v|^{1/2} \left( 1+ \sum_{n=1}^7 b_n(\mu,{\bf k})|v^{-n/3}| \right) \right] \ , 
  \end{split}
\end{equation}
with the subleading correction coefficients $a_n$, $b_n$ being rational functions of $\mu$ and ${\bf k}$. Although these expressions were determined analytically, their exact form has been omitted due to their length. Instead, the reader is referred to Appendix B of \cite{Parvizi:2021ekr} for full details.

Remarkably, the leading term of $e^{\pm}_{\mu,{\bf k}}$ (obtained by setting all $a_n$ and $b_n$ to zero) corresponds to the eigenstates of the geometrodynamics (Wheeler-DeWitt, WDW) analog of the considered background model. In fact, these eigenstates form an orthonormal basis for the Hilbert space associated with that analog (see \cite{Parvizi:2021ekr} for further details). The relation between the inner product in $\mathcal{H}_{\rm gr}$ and its Wheeler--DeWitt (WDW) analog --where in the latter case the inner product is given by a Lebesgue integral with measure ${\rm d}v$--connects the proportionality constant $C$ in \eqref{eq:xi-conv} with the norm of $\xi^{\mu}_{\bf k}$ (see Appendix~A2 of \cite{Kaminski:2010yz}). Specifically, for two eigenfunctions $\xi^{\mu}_{\bf k}$ and $\xi^{\mu'}_{\bf k}$ with proportionality constants $C$ and $C'$ respectively, their inner product is given by
\begin{equation}
  (\xi^{\mu}_{\bf k}|\xi^{\mu'}_{\bf k}) \simeq 8CC'\delta(\mu-\mu') \ .
\end{equation}
In actual computations, $C$ has been identified by methods used already in \cite{MenaMarugan:2011me,Parvizi:2021ekr}. The coefficient $C$ is determined by evaluating the interpolated extrema of $\xi^{\mu}_{\bf k}$ using polynomial interpolation. These extrema are located near the values of $v_n = v_0 \cdot 2^n$, where $v_0$ is sufficiently large for $\xi^{\mu}_{\bf k}$ to be well approximated by the combination of functions on the right-hand side of \eqref{eq:xi-conv}. The approximation of $C$ at each of these points is extracted from this approximation, and the final value of $C$ is obtained by taking the limit of these approximate values as $(1/v) \to 0$ via polynomial extrapolation.

Once the basis functions were known, the wave function profiles $\tilde{\psi}_{\bf k}(v,T)$ can be computed using equation \eqref{BO1}, where in actual calculations the Gaussian spectral profiles $c_{\bf k}(\mu)$
\begin{equation}\label{eq:app-c-gauss}
    c_{{\bf k}}(\mu) = \frac{1}{\sqrt{\pi\sigma}}e^{-\frac{(\mu-\mu_o)^2}{\sigma^2}} \, ,
\end{equation}
were used. The Romberg method was employed to evaluate the integral. Furthermore, due to the localization of the spectral profiles, the domain of the integral has been restricted to $[\mu_o-d\sigma,\mu_o+d\sigma]$ where $d$ varied from $5$ to $10$. 

Finally, the expectation values of operators $\widehat{V}^\alpha$ have been evaluated via a direct summation
\begin{equation}
  \langle \widehat{V}^\alpha \rangle(T) = \sum_{v\in 4\mathbb{Z}^+} (\alpha_o|v|)^{\alpha} |\psi(v,T)|^2 \ ,
\end{equation}
where the domain of summation can be restricted to an appropriate finite range owing to the localization of the profiles $\tilde{\psi}_{\bf k}(v,T)$. The variances have also been determined using the relation $\langle \Delta \widehat{V}^\alpha \rangle^2 = \langle \widehat{V}^{2\alpha} \rangle - \langle \widehat{V}^\alpha \rangle^2 $.

\section{Effective dynamics}
\label{sec:eff}

In this section, we derive the effective dynamics of expectation value of the volume operator with backreaction term included. To present the method, let us first consider the case of a Schrodinger-like quantization, where we have the canonical pair of operators forming the Heisenberg algebra $[\hat{v},\hat{b}] = -2i$\footnote{Note the absence of $\hbar$ on the right-hand side of the equation, which is done for convenience manifesting itself in LQC framework.}. In such a case, using Ehrenfest type equation, we have :
\begin{equation}
	\dfrac{d\langle\hat{v}\rangle}{dT} = \dfrac{\langle\big[\hat{v}, \mathbb{H}(\hat{v}, \hat{b}) \big]\rangle}{i\hslash} \simeq \dfrac{\partial {H}_{\rm gr}(\langle\hat{v}\rangle, \langle\hat{b}\rangle)}{\partial \langle\hat{b}\rangle}\; \{v, b\}+ \cdots \ ,
\end{equation}
where the higher order quantum corrections are ignored.
In polymer quantization, the operator $\hat{b}$ does not exist, and for the system at hand, the fundamental algebra is formed by the triad of symmetric operators $\widehat{\mathcal{N}}_{+} := \widehat{\mathcal{N}}^2 + \widehat{\mathcal{N}}^{-2}$ and $\widehat{\mathcal{N}}_{-} := i (\widehat{\mathcal{N}}^2 - \widehat{\mathcal{N}}^{-2} )$, with the algebra structure following the commutator $ [\hat{v},\widehat{\mathcal{N}}] = \widehat{\mathcal{N}} $. 
The gravitational Hamiltonian \eqref{Ham_grav} expressed in terms of these operators takes the form
\begin{equation}
	\widehat{H}_{\rm gr}  \ =\ - \frac{3\pi G}{8\alpha_o} \ \sqrt{|\hat{v}|}\ \big( \widehat{\mathcal{N}}_{-} \big)^2 \ \sqrt{|\hat{v}|}\ .
\end{equation} 
From there, by replacing fundamental (component) operators with their expectation values, one can write down its effective form 
\begin{equation}\label{eq:Hgr_exp}
\langle\widehat{H}_{\rm gr}\rangle  \ =\ - \frac{3\pi G}{8\alpha_o} \ \sqrt{|{v}|}\  \langle  \widehat{\mathcal{N}}_{-}\rangle^2 \ \sqrt{|{v}|}\,.
\end{equation} 
Subsequently, the (effective) equation of motion for the variable $\hat{v}$ then yields
\begin{equation}\label{eq:eom-v-eff}
	\dfrac{d\langle\hat{v}\rangle}{dT}  \ =\   \ \frac{3\pi G}{2\alpha_o}\ \langle  \widehat{\mathcal{N}}_{-}\rangle \ \langle\widehat{\mathcal{N}}_{+}\rangle \ \langle\hat{v}\rangle.
\end{equation}
Since the operators $\widehat{\mathcal{N}}_{\pm}$ satisfy the identity $\widehat{\mathcal{N}}_+^2+\widehat{\mathcal{N}}_-^2 = 4\id$, replacing $\widehat{N}_{\pm}$ with their expectation values leads to the following relation
\begin{equation}\label{eq:N-identity}
	\langle\widehat{\mathcal{N}}_{+}\rangle^2 = 4 - \langle \widehat{\mathcal{N}}_{-}\rangle^2 ,
\end{equation}
true at the effective level.
In the isotropic setting with inhomogeneities \eqref{e-m-eigenE}, 
the expectation value of the Hamiltonian constraint \eqref{trueH-Q} can be expressed as $-\langle \hat{p}_T\rangle = \langle \hat{H}_{\rm gr} \rangle + \langle {e}^n_\mathbf{k}(\hat{v}) \rangle $.
The operator ${e}^n_\mathbf{k}(\hat{v})$ is a composite operator depending, in particular, on $\hat{v}$ but not on $\widehat{\mathcal{N}}_{\pm}$; thus, it does not contribute to the equation of motion for $\langle \hat{v}\rangle$. We can read $\langle \widehat{\mathcal{N}}_{-} \rangle^2$ from the constraint (where operators are replaced by their expectation values), which yields the relation
\begin{equation}\label{eq:N}
	\langle \widehat{\mathcal{N}}_{-} \rangle^2 = \dfrac{8\alpha_o}{3\pi G}\ \frac{1}{\langle\hat{v}\rangle}  \Big( p_T + \frac{\beta \hbar k}{\ell^2\alpha_o^{\frac{1}{3}} \langle\hat{v}\rangle^{\frac{1}{3}}} \Big)  ,
\end{equation}
where $\beta = (n + 1/2)$. By substituting the above equation together with \eqref{eq:N-identity} 
into the squared form of \eqref{eq:eom-v-eff}, we obtain
\begin{equation}\begin{split}
	\Big( \dfrac{d\langle\hat{v}\rangle}{dT} \Big)^2  
    &=\  \ \Big( \frac{3\pi G}{2\alpha_o}\ \Big)^2 \langle \widehat{\mathcal{N}}_{-}\rangle^2 \ \big[4 - \langle \widehat{\mathcal{N}}_{-}\rangle^2 \big] \ \langle\hat{v}\rangle^2 \\
    &=\ \frac{6\pi G}{\alpha_o}\ \frac{1}{\langle\hat{v}\rangle}  \Big( p_T + \frac{\beta \hbar k}{\ell^2\alpha_o^{\frac{1}{3}} \langle\hat{v}\rangle^{\frac{1}{3}}} \Big) \\
    &\times \Big[4-\dfrac{8\alpha_o}{3\pi G}\ \frac{1}{\langle\hat{v}\rangle}  \Big( p_T + \frac{\beta \hbar k}{\ell^2\alpha_o^{\frac{1}{3}} \langle\hat{v}\rangle^{\frac{1}{3}}} \Big) \Big] \ \langle\hat{v}\rangle^2 , 
\end{split}\end{equation}
or in a slightly more simplified form
\begin{equation}\begin{split}
    \Big( \dfrac{d\langle\widehat{V}\rangle}{dT} \Big)^2  
    &=\ \frac{24\pi G}{\langle\widehat{V}\rangle}  \Big( p_T + \frac{\beta \hbar k}{\ell^2 \langle\widehat{V}\rangle^{\frac{1}{3}}} \Big) \\ 
    &\times \Big[1-\dfrac{2\alpha_o^2}{3\pi G}\ \frac{1}{\langle\widehat{V}\rangle}  \Big( p_T + \frac{\beta \hbar k}{\ell^2 \langle\widehat{V}\rangle^{\frac{1}{3}}} \Big) \Big] \ \langle\widehat{V}\rangle^2 , 
\end{split}\end{equation}
where, in order to write the last expression in terms of meaningful physical operators, we used the relation $\widehat{V} = \alpha_0 \hat{v}$. 
To connect the result with standard cosmology, we relate the physical volume to the scale factor as $\langle \widehat{V} \rangle = \ell^3 \langle \hat{a} \rangle^3$. In particular, this relation allows us to rewrite the energy density of the dust field as $\varrho_{\text{dust}} = p_T / \ell^3 \langle \hat{a} \rangle^3$. This, in turn, yields the effective Friedmann equation in terms of the dust energy density with backreaction correction,
\begin{equation}\label{eq:LQC-Eff-br}
    \begin{split}
	\Big( \dfrac{\langle{\dot{\hat{a}}}\rangle}{\langle\hat{a}\rangle} \Big)^2  
    &=\ \frac{8\pi G}{3 }\ \Big( \varrho_{dust} + \frac{\beta \hbar k}{\ell^6 \langle\hat{a}\rangle^{4}}  \Big)\\ 
    &\times \Big[1-\dfrac{2\alpha_o^2}{3\pi G }\ \Big( \varrho_{dust} + \frac{\beta \hbar k }{\ell^6 \langle\hat{a}\rangle^{4}} \Big) \Big],
    \end{split}
\end{equation}
where dot denotes differentiation with respect to the internal time $T$. 

For comparison, we also provide the analogous result in geometrodynamics (evaluated by repeating the same algorithm)
\begin{equation}
\Big( \dfrac{d\langle\widehat{V}\rangle}{dT} \Big)^2  =\ 24 \pi G \ \Big( p_T + \frac{\beta \hbar k}{\ell^2 \langle\widehat{V}\rangle^{\frac{1}{3}}} \Big)  \ \langle\widehat{V}\rangle . \label{eq:WDW-Eff-br}
\end{equation} 

\section{Light Propagation}
\label{sec:dispersion}

We have seen that when the energy of photons is high, due to the backreaction effects predicted  by the Born-Oppenheimer approximation, the effective background probed by them becomes frequency dependent. 
In the perturbed dressed background $\tilde{g}_{ab} dx^adx^b$ with components \eqref{dressed-metric-eq2-BO2},
the on-shell relation for the photons becomes $\tilde{g}^{ab}k_a k_b  = 0$,  so that
\begin{equation}\begin{split}
\tilde{\omega}(k) &:=  k_0\ =\  \dfrac{\tilde{N}_T}{\tilde{a}} k =: \mathcal{F}(k, T)\;\bar{\omega}, \quad \quad
\label{dispersion-BO}
\end{split}\end{equation}
where we defined
\begin{equation}\label{eq:calF-quant}
	\mathcal{F}(k, T) = \Big( \frac{\langle\hat{v^{-1}}\rangle_{\bf k} \; \langle\hat{v^{1/3}}\rangle_{\bf k}}{\langle\hat{v^{-1}}\rangle_0 \; \langle\hat{v^{1/3}}\rangle_0} \Big)^{(\frac{1}{2})}.
\end{equation} 
This indicates that the temporal frequency of the e-m modes is modified by a  mode-dependent function,  $\mathcal{F}(k, T)\neq1$. For the unperturbed background $\bar{g}_{ab} dx^adx^b$ with components \eqref{dressed-metric-eq2a}, we have $\bar{\omega} = \bar{N}_T / \bar{a}\, k $.

After finding $\langle\widehat{V}^n\rangle_{(0, {\bf k})}$ for $n = 1, -1, 1/3$ using LQC dynamics \eqref{BO1} or geometrodynamics \eqref{eq:WDW-Eff-br}, we can compute the modification function $\mathcal{F}(k, T)$ to the dispersion relation in four different regimes for both perturbed and unperturbed cases:

Regime A : {\it genuine quantum LQC framework,} where $\mathcal{F}$ has been evaluated via full expression defined in \eqref{eq:calF-quant}, with the expectation values determined numerically from a class of semiclassical quantum states constructed in Eq.~\eqref{BO1};\\
Regime B: {\it hybrid LQC framework,} where the approximation $\langle\widehat{V}^n\rangle \approx \langle\widehat{V}\rangle^n$ is employed and the higher order correction in the semi-classical expansion of different powers of the volume \cite{Bojowald:2005cw} in terms of the central Hamburger moments, namely, 
\begin{equation}
	\langle \widehat{V}^{\alpha} \rangle = \langle \widehat{V} \rangle^{\alpha} + \sum_{i=1}^\infty    \left({\begin{array}{c}
		\alpha \\
		i  \\
		\end{array} } \right) \langle \widehat{V} \rangle^{\alpha-i} G^{i00} \, ,
\end{equation}
where $G^{i00}=\langle (\delta \widehat{V})^i \rangle$, are ignored. This approximation results in the simplified formula 
\begin{equation}
    \mathcal{F}(k, T) = \Big( \langle\hat{v}\rangle_o / \langle\hat{v}\rangle_{\bf k} \Big)^{1/3}  \, .
\end{equation}
Regime C: {\it effective LQC framework,} where we do not compute the expectation value of operators with respect to a quantum state; instead, we use the classical trajectories for geometry variables as determined via Eq.~\eqref{eq:LQC-Eff-br}, which includes LQC quantum correction terms related to the parameter $\alpha_0$. Then we use their values to  evaluate $\mathcal{F}(k, T)$, using
\begin{equation}\label{F-effective}
    \mathcal{F}(k, T) = \frac{a_0(T)}{a(T)} .
\end{equation}
Regime D: {\it effective geometrodynamics framework,} where the effective dynamics approach is applied to the model with geometry quantized using the geometrodynamics framework. It yields Eq.~\eqref{eq:WDW-Eff-br} at the semi-classical limit, which effectively reproduces general relativity (GR) dynamics asymptotically. This equation is then used to evaluate $\mathcal{F}(k, T)$. 

In order to obtain a comprehensive picture of the possible modifications, the function $\mathcal{F}$ has been investigated in all these four regimes.

The genuine quantum studies, while the most accurate, are limited, as they can only be performed for a finite set of examples. In these cases, states with Gaussian spectral profiles $c_{\mathbf{k}}(\mu)$ (see Eq.~\eqref{eq:app-c-gauss}) are selected, where the parameters $\mu_0$ and $\sigma$ are related to the expectation value and variance of the clock-field momentum as follows
\begin{equation}
    \langle p_T \rangle = \sqrt{12\pi G}\hbar\mu_o , \qquad \langle\Delta p_T\rangle = \sqrt{12\pi G}\hbar\sigma \ .
\end{equation}
We select the same profile for both the backreacted and background states. In brief, this choice is motivated by (i) the fact that in both cases the eigenspaces are one-dimensional, allowing the energy eigenbasis to be chosen as real eigenfunctions, each exhibiting the qualitative properties of a reflected wave, and (ii) the fact that the differences in the asymptotics of the background and backreacted states agree, up to a phase shift 
between the asymptotic wave components, at leading order.\\
Given the spectral profiles \eqref{eq:app-c-gauss}, both the backreacted and background states have been integrated out using \eqref{BO1} and its analog, where $\xi_{\rm K}^{\mu}$ is replaced by the eigenfunctions of $\hat{H}_{\rm gr}$, respectively. The computations were performed with the Numerical LQC library (see \cite{MenaMarugan:2011me} for technical details). Subsequently, the expectation values of observables present in \eqref{eq:calF-quant} (being in $v$-representation multiplication operators) have been evaluated on the uniform lattice in $T$. 
\section{Analytic properties of effective dynamics}\label{sec:props-DR}
All the nontrivial properties of the electromagnetic dispersion relation are captured in the function $\mathcal{F}(k,T)$ defined in \eqref{dispersion-BO}. In the fully quantum regime, whether it is the one where the spatial background is quantized by means of geometrodynamics (Wheeler-DeWitt) or Loop Quantum Cosmology, probing its behavior requires employing numerical methods quite intensively. However, the simplification of the systems provided by the methods of the (classical) 0th order effective dynamics is sufficiently radical to give hope for studying $\mathcal{F}$ in that regime analytically. Indeed, in the next subsection, we will derive the exact analytic form of $\mathcal{F}$ for the case based on WDW geometry, while for the LQC based case, the critical properties regarding its low energy limit will be shown in the subsequent subsection.
\subsection{Effective Wheeler-DeWitt case}
\label{app:anal-wdw}
Let us start with the model following from the geometrodynamic description of the spatial geometry. Our point of departure is the effective evolution equations \eqref{eq:WDW-Eff-br}, where the background (unperturbed) spacetime case corresponds to $\beta=0$. The solutions to them will, in turn, determine the function $\mathcal{F}$ via \eqref{F-effective}. \\
Let us first find the solutions to \eqref{eq:WDW-Eff-br}. For that, it is convenient to rewrite it in terms of the variable $x$ defined by $V = x^{-3}$
\begin{equation}
    \dot{x}^2 = \frac{8\pi G}{3} x^5 (p_T + \ell^{-2} \beta\hbar k x) \ .
\end{equation}
As this is a first-order homogeneous equation, the implicit form of the solution can be expressed as a quadrature, which may subsequently be integrated analytically. This yields
\begin{equation}\label{eq:tx_sol_implicit}\begin{split}
    T(x) - T_o &= \pm \int_x^{\infty} \frac{\rd x'}{\sqrt{\frac{8\pi G}{3} {x'}^5 (p_T +\ell^{-2} \beta\hbar k x')}}  \\
    &= \mp \Big( \frac{\sqrt{p_T+ \ell^{-2} \beta\hbar k x}(2 \ell^{-2} \beta\hbar k x-p_T)}{p_T^2\sqrt{6\pi G} x^{\frac{3}{2}} } 
      \\
      &- \frac{2(\ell^{-2} \beta\hbar k)^{\frac{3}{2}}}{p_T^2\sqrt{6\pi G}}\Big) \, ,
\end{split}\end{equation}
where $T_o$ is the time at which the trajectory reaches the cosmological singularity. In terms of variables $x$, the function $\mathcal{F}(k,T)$ takes a very simple form
\begin{equation}\label{eq:F-xxo}
    \mathcal{F}(k,T) = \frac{x(T)}{x_o(T)} \ ,
\end{equation}
where $x_o(T)$ corresponds to the solution to \eqref{eq:tx_sol_implicit} with $\beta=0$. Determining it uniquely requires, however, synchronizing $T_o$ for both background and backreacted trajectory. For that, we follow the reasoning already applied to the case of the quantum geometry presented in Sec.~\ref{quantum-sys} in LQC framework. At the quantum level, the WDW counterpart $\widehat{\ub{H}}_{\rm gr}$ of the gravitational Hamiltonian $\widehat{H}_{\rm gr}$ admits a $U(1)$ family of self-adjoint extensions (see \cite{Husain:2011tm}). Each extension features a nondegenerate spectrum, with energy eigenstates corresponding to reflected plane waves subject to extension-dependent reflective boundary conditions at $V=0$. This structure allows the energy eigenbasis to be chosen as real. The WDW counterpart of the backreacted Hamiltonian (the operator appearing on the left-hand side of \eqref{eq:H-corr}) exhibits the same features, which allows the backreacted energy eigenbasis to be fixed in the same way (for the explicit form of $\widehat{\ub{H}}_{\rm gr}$ in WDW approach, see \cite{Parvizi:2021ekr}). As a consequence, we can set $T_o$ for both trajectories to be equal and subsequently fix the time translation freedom by setting them to $T_o=0$. Taking the solution for $\beta = 0$ from \eqref{eq:tx_sol_implicit},
\begin{equation}
    T = \frac{\pm 1}{\sqrt{6\pi G p_T}\, x_o^{3/2}} \ ,
\end{equation}
we invert this relation and substitute it into \eqref{eq:F-xxo}, obtaining
\begin{equation}\label{F-WDW-eff-app}\begin{split}
    \mathcal{F}^{\frac{3}{2}}(k,x) 
    &= \frac{x^{\frac{3}{2}}}{x^{\frac{3}{2}}_o(T(x))}  \\
    &= p_T^{-\frac{3}{2}}
      \Big( \sqrt{p_T+ \ell^{-2} \beta\hbar kx} (p_T-2 \ell^{-2} \beta\hbar kx) \\
      &+ 2(\ell^{-2}\beta\hbar k)^{\frac{3}{2}} x^{\frac{3}{2}} \Big) \ .
\end{split}\end{equation}
From that, one immediately observes that
\begin{equation}\label{eq:wdw-limit-app}
    \lim_{x \to 0} \mathcal{F}(k,x) = 1 \ ,
\end{equation}
so the model reproduces general relativity exactly in the low-energy limit, 
without any modification to the speed of light.

\subsection{Effective loop quantum cosmology case}
\label{app:anal-LQC}

For the effective LQC model, we can repeat (with a few modifications) the procedure described in the previous subsection. In this case, our starting point is the equation \eqref{eq:LQC-Eff-br}, where, again, the background (unperturbed) spacetime evolution is captured by the case $\beta=0$. Rewriting it with respect to $x:=V^{-\frac{1}{3}}$, we get
\begin{equation}
    \dot{x}^2 = \frac{8\pi G}{3} x^5 (p_T+ \ell^{-2} \beta\hbar kx)\left(1-\frac{2\alpha_o^2}{3\pi G}x^3(p_T+\ell^{-2}\beta\hbar kx)\right) \ . 
\end{equation}
The solution can be written in an implicit form
\begin{equation}\begin{split}\label{eq:tx-lqc}
    &T(x) - T_o = \\
    &\pm \int_x^{x_o} \frac{\sqrt{\frac{3}{8\pi G}}{x'}^{-\frac{5}{2}}\rd x'}{\sqrt{(p_T+\ell^{-2}\beta\hbar kx)\left(1-\frac{2\alpha_o^2}{3\pi G}x^3(p_T+\ell^{-2}\beta\hbar kx)\right)}} \ ,
    \end{split}
\end{equation}
where $T_o$ is the time of the bounce and $x_o$ (inverse scale factor at the bounce) is the solution to the equation
\begin{equation}\label{eq:x0}
   \ell^{-2} \beta\hbar k x^4 + p_T x^3 = \frac{3\pi G}{2\alpha_o^2} \ ,
\end{equation}
satisfying the condition
\begin{equation}
    0 < x < \left( \frac{3\pi G}{2\alpha_o^2p_T} \right) ^{\frac{1}{3}} \ .
\end{equation}
The right hand side of \eqref{eq:tx-lqc} can be evaluated for $\beta=0$ analytically, giving
\begin{equation}\label{eq:x0T}
    x_o^{-\frac{3}{2}} = p_T^{\frac{1}{2}} \sqrt{6\pi G (T-T_o)^2  + \frac{2\alpha_o^2}{3\pi G}} \ .
\end{equation}
upon substitution $\ell x = 1/a$, we obtain
\begin{equation}
    \ell a(t) = p_T^{\frac{1}{3}} \left( 6\pi G(T-T_o)^2 + \frac{2\alpha_o^2}{3\pi G} \right)^{\frac{1}{3}} \ .
\end{equation}
Combining the above equation with \eqref{eq:tx-lqc} and setting the same $T_o$ for both the background and backreacted cases, one can write down the expression for $\mathcal{F}$ as a function of $x$ in integral form. In order to probe its behavior in the low energy limit, we need to check its properties near $x=0$. By expanding the integrand in \eqref{eq:tx-lqc}, we note that
\begin{equation}
    T-T_o = \frac{2}{3} x^{-\frac{3}{2}} \left( \sqrt{\frac{3}{8\pi G p_T}} + O(x) \right) \ ,
\end{equation}
which in turn yields
\begin{equation}\label{eq:LQC-limit-app}
    \mathcal{F}(x) = 1 + O(x) \ .
\end{equation}
As a consequence, the GR propagation is restored in the low energy limit as in the geometrodynamics case.

The results obtained within the genuine quantum framework have been subsequently compared against the semiclassical effective description for both LQC and Geometrodynamics. In the former case, the background and backreacted trajectories have been found analytically, where we assumed $\langle \hat{V} \rangle \approx \ell^3 a^3(T)$ and $\langle \hat{a} \rangle \approx a(T)$.  
Once the classical trajectory of backreacted $a(T)$ and background $a_o(T)$ is found, the effective dispersion relation coefficient ${\cal F}$ is determined via eq.~\eqref{F-effective}. In practical calculations of the backreacted effective trajectories in LQC, rather than employing the integral form, we directly solved the initial 
value problem originally defined by \eqref{eq:LQC-Eff-br}, which was subsequently reformulated as a globally regular second-order initial value problem, as specified in Appendix~\ref{app:eff-ivp}.
\section{The results and conclusions}
\label{sec:results}
In actual evaluations, the dispersion relation coefficient $\mathcal{F}$ has been determined in four regimes.
While regimes \textit{A - B} allow one to test the accuracy of simplifications and determine the origins of particular effects, regime \textit{C} provides a point of reference in distinguishing the effects of loop representation in comparison with the standard quantum mechanical treatment (geometrodynamics).
All approaches have been applied to a population of states peaked around 
$p_T$ in the range $500$ to $5 \times 10^3 \hbar/\sqrt{G}$, with relative 
variance $\Delta p_T/p_T \in [0.05,0.1]$. The number of electromagnetic 
particles considered varies from $1$ to $10$, and the mode index ranges 
from $25$ to $125$.
\begin{figure*}[tbh!]
    \begin{center}
    \subfloat[]{\includegraphics[width=.45\textwidth]{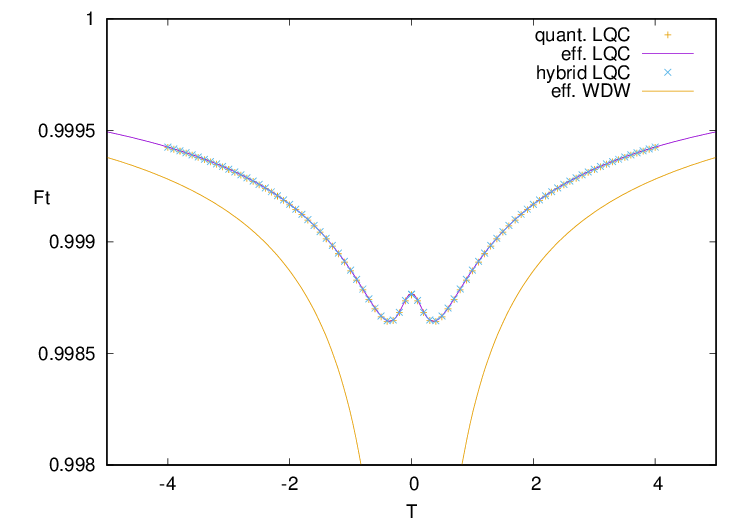}\label{fig:Ft-zoom}}
    \hspace{0.1cm}
    \subfloat[]{\includegraphics[width=0.45\textwidth]{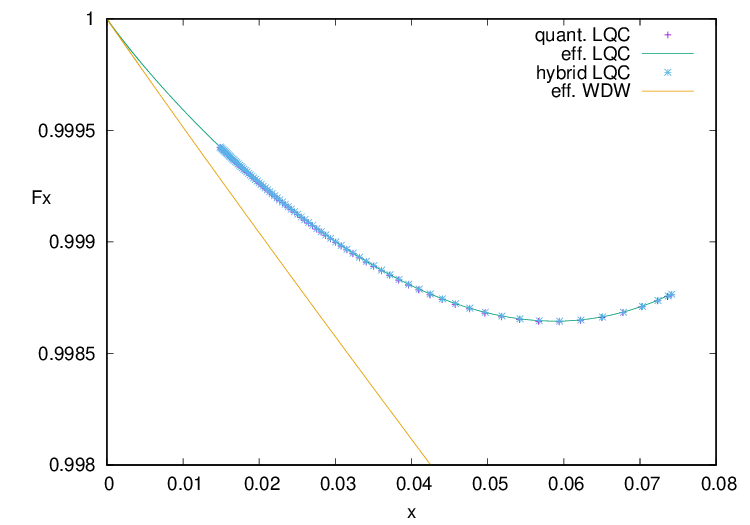}\label{fig:Fx-main}}
    \end{center}
    \caption{Coefficient $\mathcal{F}$ evaluated via methods \textit{(A -- D)} for a Gaussian state peaked about $p_T=10^3\hbar/\sqrt{G}$ with relative variation $\Delta p_T/p_T = 0.05$ with a single particle mode of $k=100$ plotted as function of dust time \eqref{fig:Ft-zoom} and $x=V^{-1/3}$ \eqref{fig:Fx-main}. One sees that $(i)$ the dispersion relation of GR is restored in low energy limit ($\mathcal{F} \to 1$), and $(ii)$ the modifications in the loop approach are smaller than those in geometrodynamics.}
    \label{fig:F-main}
\end{figure*}

\begin{figure}[tbh!]
    \begin{center}
    \includegraphics[width=.45\textwidth]{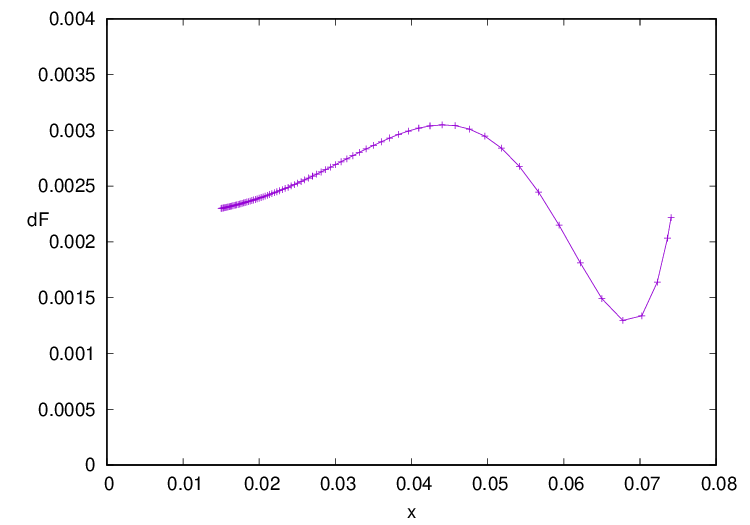}
    \end{center}
    \caption{Relative difference between the dispersion‑relation coefficients \(\mathcal{F}\) obtained from the genuine quantum treatment (Regime A) and the hybrid LQC method (Regime B), evaluated for the configuration shown in Fig.~\ref{fig:F-main} One can see a regular behavior indicating a stabilization as $x\to 0$. }
    \label{fig:dFx-LQC}
\end{figure}

A representative example of our findings is shown in Fig.~\ref{fig:F-main}. Across all cases analyzed, several robust features consistently emerge. First, the results obtained from the genuine quantum treatment (Regime A - Purple curve) display very small deviations from those of the hybrid LQC method (Regime B - Blue curve) for the coefficient \(\mathcal{F}\). These differences yield results that are essentially indistinguishable from the effective one (Regime C - Green curve).
A second general feature concerns the relative modification of the dispersion relation, quantified by  
\begin{equation}
\delta\mathcal{F}(T)=\frac{|\mathcal{F}_{q}(T)-\mathcal{F}_{h}(T)|}{1-\mathcal{F}_{h}(T)} ,
\end{equation}
where \(\mathcal{F}_{q}\) and \(\mathcal{F}_{h}\) denote the outcomes of the genuine quantum and hybrid approaches, respectively. Within the domain accessible to numerical analysis (see Fig.~\ref{fig:dFx-LQC}), this quantity remains bounded. Moreover, the behavior of \(\delta\mathcal{F}\) in the probed region strongly suggests that the boundedness persists all the way to the low‑energy limit \(x\to 0\).
A further universal observation is that the LQC‑derived coefficient \(\mathcal{F}\) always satisfies  
\begin{equation}
\mathcal{F}_{\rm WDW} < \mathcal{F}_{\rm LQC} < 1 ,
\end{equation}
throughout the entire domain of applicability and for all methods considered (Regimes A–D). Consequently, the LQC corrections to the dispersion relation are systematically smaller than those predicted by the geometrodynamical (WDW) framework.
Interestingly, the behavior of \(\mathcal{F}_{\rm LQC}\) near the bounce differs from expectations based on the WDW case. Whereas \(\mathcal{F}_{\rm WDW}\) exhibits increasing deviations from unity as the matter energy density grows (see Fig.~\ref{fig:Ft-zoom}), the LQC coefficient does not reach its maximum deviation at the bounce point \(T=0,\; x\approx 0.074\). Instead, in the strongly quantum regime surrounding the bounce, the departure of \(\mathcal{F}_{\rm LQC}\) from unity actually decreases as the energy density increases.
Finally, both LQC and geometrodynamics share an exact low‑energy limit: the effective descriptions (Regime C and Regime D) yield  
\begin{equation}
\lim_{x\to 0}\mathcal{F}_{\rm LQC}(x)
=\lim_{x\to 0}\mathcal{F}_{\rm WDW}(x)
=1 ,
\end{equation}
a result confirmed analytically in Eqs.~\eqref{eq:wdw-limit-app} and \eqref{eq:LQC-limit-app}. This establishes that both frameworks recover the standard dispersion relation in the infrared regime.

On the level of the 0th order effective dynamics, all the results are exact, while at the genuine quantum level, they have been established numerically within a specified domain. However, the relative differences in the dispersion relation correction, $(1 - \mathcal{F}(a))$, between the genuine quantum and effective approaches are bounded within the probed domain (Fig.~\ref{fig:dFx-LQC}). Their behavior within this domain, together with experience from other models in LQC, provides strong justification for extrapolating this bound to low energies ($x \to 0$) with a high level of confidence. Thus, there are strong arguments that the results established at the effective level also hold at the genuine quantum level \cite{Ashtekar:2006rx, Ashtekar:2006wn}, although a full numerical exploration of the genuine quantum regime at low energies is computationally expensive.

\begin{figure*}[tbh!]
    \begin{center}
    \subfloat[]{\includegraphics[width=.45\textwidth]{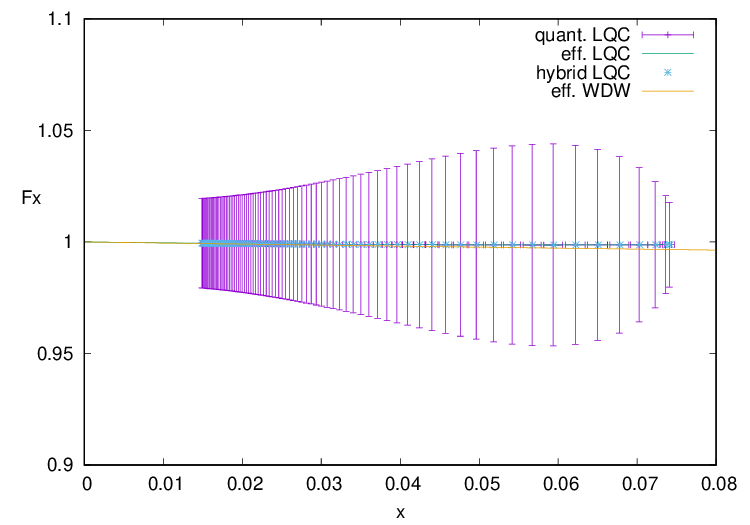}\label{fig:Fx-err-real}}
    \hspace{0.1cm}
    \subfloat[]{\includegraphics[width=0.45\textwidth]{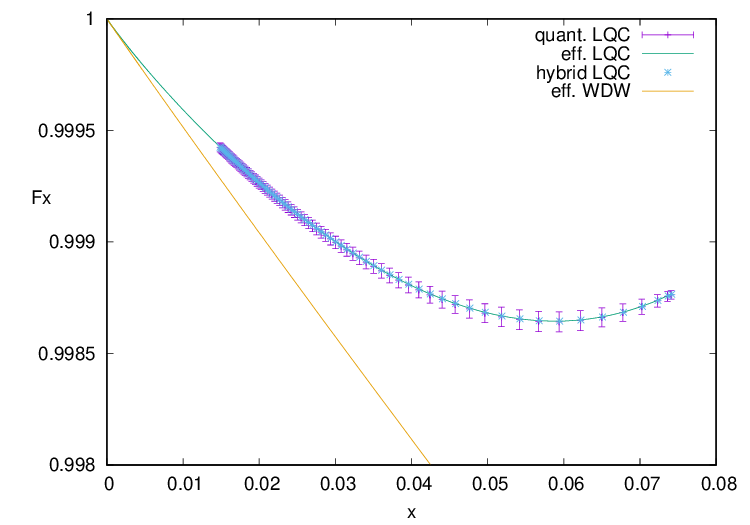}\label{fig:Fx-err-extr}}
    \end{center}
    \caption{Estimate of the uncertainty in \(\mathcal{F}(x)\) for the configuration shown in Fig.~\ref{fig:F-main} within the LQC framework (a), together with its extrapolation to a state whose spectral profile is rescaled by a factor of \(10^{3}\) in \(k\) and whose particle number \(\beta\) is increased by \(10^{4}\) (b). This scaling corresponds to shifting the peak of \(p_{T}\) to a value \(10^{3}\) times larger while keeping the radiation energy density fixed. Although in $(a)$ the effects of quantum gravity are masked by the uncertainty, in $(b)$ these effects are clearly visible.}
    \label{fig:Fx-err}
\end{figure*}

The above results demonstrate specific modifications to the dispersion relation relative to that of general relativity at high energies, for both geometrodynamics and LQC-based models. However, it must be remembered that the dispersion relation coefficient is determined by expectation values 
of quantum operators, and therefore carries an intrinsic uncertainty due to quantum variance. One should thus ask whether the observed corrections are indeed large enough to be significant. In order to verify that, a variance of $\mathcal{F}$ has been derived as the standard quadratic deviation for a function of observables (following a standard technique used to estimate the errors of composite quantities due to measurement errors). An example of the result is presented in Fig.~\ref{fig:Fx-err-real}. As one can see, for the cases actually evaluated numerically, the answer is in the negative: the deviations completely mask the studied effects of quantum gravity. However, upon rescalling of the state spectral profiles \eqref{eq:app-c-gauss} via 
\begin{equation}\label{eq:spectr-scale}
    \mu \mapsto C\mu , \qquad \sigma\mapsto C\sigma
\end{equation}
(which does not change the relative dispersion in $p_T$) the observables entering $\mathcal{F}$ (and their variances) scale approximately as follows
\begin{subequations}
\begin{align}
    \langle V^{\frac{1}{3}} \rangle &\mapsto C^{\frac{1}{3}} \langle V^{\frac{1}{3}} \rangle , &
    \langle \Delta V^{\frac{1}{3}} \rangle 
        &\mapsto C^{-\frac{2}{3}} \langle \Delta V^{\frac{1}{3}} \rangle , \\
    \langle V^{-1} \rangle &\mapsto C^{-1} \langle V^{-1} \rangle , &
    \langle \Delta V^{-1} \rangle &\mapsto C^{-2} \langle \Delta V^{-1} \rangle . &
\end{align}
\end{subequations}
Furthermore, as discussed in Appendix.~\ref{app:scaling}, simultaneous scaling of the particle number $\beta\mapsto C^{\frac{4}{3}}\beta$ (preserving the radiation energy density at the same time with respect to the bounce) leaves the dispersion relation coefficient $\mathcal{F}$ invariant \eqref{eq:Bk} and \eqref{eq:LQC-scaling}. This allows us to extrapolate the uncertainty of $\mathcal{F}$ to scenarios in which the dust momentum of the universe is increased by a factor of $C$, while its relative dispersion remains unchanged. An example for $C=10^3$ is shown in Fig.~\ref{fig:Fx-err-extr}. One can see that the uncertainty is now well below the deviations of $\mathcal{F}$ from GR. Unfortunately, performing numerical simulations of states with $p_T$ values that high is, at present, beyond our technical capabilities.
In this work, we investigate in detail the possible modification to massless matter propagation through quantum spacetime as a possible effect of the quantum nature of spacetime. To this end, we applied the so-called dressed metric approach \cite{Ashtekar:2009mb} to quantum gravity, following an approach inspired by the Born--Oppenheimer approximation \cite{Giesel:2009at}. As a propagating matter, we selected the electromagnetic field emulated by a triad of inhomogeneous scalar fields, which, upon decomposition onto modes, form a hybrid model of (now homogeneous) matter degrees of freedom living on an isotropic spacetime. This particular model was already considered in \cite{Lewandowski:2017cvz}. However, due to the assumptions adopted there, the results could not be regarded as definitive, particularly leading to the troubling conclusion of superluminal light propagation at low energies. The technique employed in that work was subsequently refined in the context of Oppenheimer--Snyder collapse in \cite{Parvizi:2021ekr}. In the present study, we apply these refinements together with a careful numerical analysis to reexamine the predictions of \cite{Lewandowski:2017cvz}.
Previous theoretical studies (see for example, \cite{Gambini:1998it, Alfaro:1999wd, Alfaro:2001rb}) on the modification to the dispersion relation have often relied on ad hoc assumptions or remained confined to the semi-classical regime. In contrast, our aim was to establish a more robust theoretical and numerical framework to address the chromatic dispersion effect in quantum gravity.

The analysis leads to the following main conclusions: (i) due to backreaction effects, the dispersion relation of the electromagnetic field in quantum spacetime becomes frequency-dependent; (ii) both particle number and particle frequency enter the modified dispersion relation as amplification parameters; (iii) contrary to previously reported results \cite{Lewandowski:2017cvz}, the standard dispersion relation of general relativity is recovered in the low-energy limit; (iv) the deviations probed by this model are actually bounded and weaker in the polymeric (LQC) framework compared to those in geometrodynamics; and (v) in all cases considered, the propagation of electromagnetic radiation remains subluminal.

One must remember, however, that despite significant improvements, the results presented here cannot yet be regarded as final, since they were obtained within a simplified model. For definitive conclusions in the context of loop quantization, it is necessary to employ genuinely inhomogeneous formulations of loop quantum gravity.


\begin{acknowledgments}
    This work was supported in part by the Polish National Center for Science (Narodowe Centrum Nauki -- NCN) grant OPUS 2020/37/B/ST2/03604. The work of AP was supported by a grant of the Transilvania University of Brașov (UNITBV), offered through the Transilvania Fellowship for Postdoctoral Research/Young Researchers program.
\end{acknowledgments}

\appendix

\section{Chromatic dispersion in materials}\label{app-CDMMaterials}
The treatment presented in the main body of the paper is an adaptation of well known techniques used in probing the backreaction effects of the e-m radiation propagating through solid state matter. In particular, the chromatic dispersion effects of quantum gravity that arise in our analysis are analogous to the familiar dispersion phenomena encountered in linear optics. For this reason, we include in this Appendix a brief reminder and a direct comparison of how optical susceptibility is derived within linear response theory. It is based on the interaction between atoms and the e-m field, employing quantum-mechanical perturbation theory applied to the atomic wave function \cite{Boyd:2008eba}. In this context, the total Hamiltonian for the system is 
\begin{equation}
    \widehat{H} = \widehat{H_0} + \lambda \widehat{V} + \widehat{H}_{EM} \, ,
\end{equation}
 where $\widehat{H_0}$ is the Hamiltonian term corresponding to the atoms, $\widehat{V}$ describes the interaction of the atom(s) with the e-m field, and $\lambda$ is the perturbation parameter. 
The interaction term is taken to be of the form
\begin{equation}
    \widehat{V} = - \mathbf{\hat{\mu}} \cdot \mathbf{\hat{E}} \, ,
\end{equation}
where $\mathbf{\hat{\mu}} = - e \mathbf{\hat{r}}$ is the electric-dipole moment operator and $e$ is the electric charge. Through the use of perturbation theory, we now seek a solution to Schrödinger’s equation in the form
\begin{equation}\label{eq:mstate}
    \tilde{\psi} (\mathbf{r},t) = \psi^0 (\mathbf{r},t) + \lambda \psi^{(1)} (\mathbf{r},t) + \dots \ ,
\end{equation}
at first order in perturbation, while the states satisfy the following set of equations,
\begin{subequations}\label{eqs:L-M}
    \begin{align}
        i\hbar \frac{\partial \psi^{(0)}}{\partial t} &= \widehat{H_0} \psi^{(0)}, \label{eq:zero}\\
        i\hbar \frac{\partial \psi^{(1)}}{\partial t} &= \widehat{H_0} \psi^{(1)} + \widehat{V} \psi^{(0)}. \label{eq:first}
    \end{align}
\end{subequations}
To calculate the first order perturbation, one makes use of the fact that the solutions to the free electromagnetic Hamiltonian are composed of the plane waves $\mathbf{E}(t) = \sum_p \mathbf{E}(\omega_p) e^{-i\omega_p t }$. Given that, one can calculate and expand the matrix elements of the perturbing Hamiltonian $ \hat{V}$ as follows,
\begin{equation}
    \langle m | \widehat{V} | l \rangle = \int u^\star_m \, \widehat{V} \, u_l \, d^3r = - \sum_p \mathbf{\mu}_{ml} \cdot \mathbf{E}(\omega_p) e^{-i \omega_p t} \ ,
\end{equation}
where $\mathbf{\mu}_{ml} = \int u^\star_m \, \mathbf{\hat{\mu}} \, u_l \, d^3r$ is known as the electric-dipole transition moment and $u_l$ are the unperturbed eigenstates of the atoms. When we have the expectation value of the ``perturbation'' (the interacting term of the Hamiltonian) $\hat{V}$ with respect to the unperturbed states $\psi^{(0)}$, the state $\tilde{\psi}$ can be calculated using equation \eqref{eq:mstate}.

When the state $\tilde{\psi}$ is constructed, one can compute the linear first order correction to the susceptibility of the material. This requires evaluating the expectation value of the electric dipole moment,
$\langle \mathbf{\widehat{p}} \rangle \simeq \langle \tilde{\psi} | \mathbf{\widehat{\mu}} | \tilde{\psi} \rangle = \langle \psi^{(0)} | \mathbf{\widehat{\mu}} | \psi^{(0)} \rangle + \langle \psi^{(0)} | \mathbf{\widehat{\mu}} | \psi^{(1)} \rangle + \langle \psi^{(1)} | \mathbf{\widehat{\mu}} | \psi^{(0)} \rangle$, up to the first order correction\footnote{The first term is zero.}.
Assuming a uni-directional electric field and using the perturbed state of the atom, the general expression for the electric-dipole moment will have the following form
\begin{equation}
    \langle {\widehat{p}} \rangle = \sum_p \alpha(\omega_p) E(\omega_p) e^{- i \omega_p t} \, \, .
\end{equation}
Here, $\alpha(\omega_p)$ depends on the atomic structure and the frequency of the propagating electric field. For N atoms, the total electric dipole moment is given by $P = N \langle {\widehat{p}} \rangle$. Introducing the linear susceptibility defined by the relation: $ P = \epsilon_0 \sum_p \chi(\omega_p) E(\omega_p) e^{(-i \omega_p t)}$ \cite{Boyd:2008eba}, we can now obtain the linear \textit{frequency-dependent} electric susceptibility as follows:
\begin{equation}
    \chi(\omega_p) = \frac{N}{\epsilon_0} \, \alpha(\omega_p),
\end{equation}
for explicit relations between $\chi(\omega_p)$ and $\alpha(\omega_p)$, see \cite{solyom2008fundamentals,Boyd:2008eba}.

Since our L--G system has an additional layer of complexity--because $\widehat{H}_{EM}$ itself is a composite operator involving both geometric and field operators, and free states for the electromagnetic field are not available in the L--G system unlike in the L--M system--the procedure to solve the system here is comprehensively more complicated than in the L--M case. 
As explained in more detail in \cite{Parvizi:2021ekr} and illustrated 
diagrammatically in Fig.~\ref{fig:schem}, we first solve the unperturbed 
background geometry equations to obtain $\psi_o(v, T)$. Next, we employ the concept of a dressed metric to determine the corresponding background metric for the field. Using the metric components $\bar{N}(T)$ and $\bar{a}(T)$ of this dressed background, we construct the field states $\phi_{\mathbf{k}}(Q_{\mathbf{k}}, T)$ with eigenvalues $e^n_{\mathbf{k}}(\hat{v})$. These results are then used to derive the perturbed first-order geometric state $\tilde{\psi}(v, T)$. Finally, we obtain the corresponding mode-dependent dressed background for the field, given by $\tilde{N}(T,k)$ and $\tilde{a}(T,k)$, which are constructed from the mode-dependent geometric quantities $\langle \hat{V}^n \rangle_{\mathbf{k}}$.

\section{Scaling properties of effective $\mathcal{F}$}
\label{app:scaling}

One of the interesting properties of the effective background trajectories $a(T)$ in both WDW and LQC cases is that they both scale with $p_T^{\frac{1}{3}}$. This scaling symmetry corresponds to the freedom of selecting the compact region of the Universe (the fiducial cell) used to define momenta and Hamiltonian in cases where the Universe spatial slice is noncompact. One could expect this property to hold also for the backreacted geometries. Here we show that this is indeed the case (in both WDW and LQC cases), provided that an adequate contribution of the e-m modes is kept.

Let us start with WDW, where $\mathcal{F}$ is determined analytically as given in \eqref{F-WDW-eff-app}. By introducing the auxiliary (scaling invariant) variable 
\begin{equation}\label{eq:xy-transform}
    y := p_T^{\frac{1}{3}}\, x \ ,
\end{equation}
one can rewrite it in the following way
\begin{equation}
    \mathcal{F}^{\frac{3}{2}}(k,y) = \sqrt{1+B_ky}\,(1-2B_ky) + 2(B_ky)^{\frac{3}{2}} \ , 
\end{equation}
where 
\begin{equation}\label{eq:Bk}
    B_k := \ell^{-2} \beta\hbar k p_T^{-\frac{4}{3}} \ ,
\end{equation}
encodes the contribution of the e-m field mode to the energy relative to the clock field. 

In LQC the same transformation \eqref{eq:xy-transform} applied to \eqref{eq:tx-lqc}, \eqref{eq:x0}, \eqref{eq:x0T} again allows to write $\mathcal{F}$ in a scaling invariant (though depending on $B_k$ defined in \eqref{eq:Bk}) form
\begin{subequations}\label{eq:LQC-scaling}\begin{align}
    \mathcal{F}^{\frac{3}{2}}(k,y) &= y^{\frac{3}{2}} \sqrt{6\pi G \Delta T^2(k,y) + \frac{2\alpha_o^2}{3\pi G}} \ , \\
    \Delta T(k,y) &= \int_{y}^{y_o} \frac{{y'}^{-\frac{5}{2}} \rd y'}{\sqrt{\frac{8\pi G}{3}(1+B_ky)(1-\frac{2\alpha_o^2}{3\pi G}y^3(1+B_ky))}} \ ,
\end{align}\end{subequations}
where $y_o$ satisfies
\begin{equation}
    B_k y_o^4 + y_o^3 = \frac{3\pi G}{2\alpha_o^2} \ , \quad 
    0<y_o<\left(\frac{3\pi G}{2\alpha_o^2}\right)^{\frac{1}{3}}\ .
\end{equation}
This invariance is particularly useful in extrapolating the genuine quantum results beyond the domain currently accessible for numerical probing.
\section{Initial value problem for the effective Friedmann equation}
\label{app:eff-ivp}

While in the geometrodynamical framework the effective Friedmann equation \eqref{eq:WDW-Eff-br} can be integrated analytically, yielding a closed‑form expression for \(a(T)\), i.e., \eqref{eq:tx_sol_implicit}, the corresponding effective equation in the LQC framework \eqref{eq:LQC-Eff-br} admits no such solution when \(\beta \neq 0\). Despite extensive analysis, we were unable to obtain an analytic expression for \(a(T)\) in this case, which forced us to rely on numerical methods to solve the effective LQC equation. However, the initial value problem using \eqref{eq:LQC-Eff-br} directly is not suitable, as it violates the assumptions of the uniqueness theorem at the bounce point. In order to sidestep this issue, we employed the method used for example in \cite{Pawlowski:2011zf}, reformulating it as the 2nd order one. Indeed, by differentiating the 1st order equation following from \eqref{eq:LQC-Eff-br} 
\begin{subequations}\label{eq:ivp-lqc}\begin{align}
    \dot{a} &= \pm a \sqrt{G(a)} , \\
    \begin{split}
    G(a) &= \frac{8\pi G}{3 }\ \Big( \frac{p_T}{\ell^3a^3} + \frac{\beta \hbar k}{\ell^6 a^{4}}  \Big) \\ 
    &\times \Big[1-\dfrac{2\alpha_o^2}{3\pi G }\ \Big( \frac{p_T}{\ell^3a^3} + \frac{\beta \hbar k }{\ell^6 a^{4}} \Big) \Big] \ ,
    \end{split}
\end{align}\end{subequations}
and substituting its right hand side for $\dot{a}$ we get
\begin{equation}
    \ddot{a} = a G(a) + \frac{1}{2} a^2 \frac{\rd G(a)}{\rd a} \ ,
\end{equation}
which possesses all the regularity conditions required for the uniqueness of the solution. In principle, now the problem (being the 2nd order one) requires $\dot{a}$ at the initial time. This, however, can be provided via \eqref{eq:ivp-lqc} (with sign selected appropriately for the expanding/contracting epoch) outside of the bounce point or the set
\begin{equation}
    \dot{a}(T_o) = 0 \ , \qquad 
    \dfrac{2\alpha_o^2}{3\pi G }\ \Big( \frac{p_T}{\ell^3a(T_o)^3} + \frac{\beta \hbar k }{\ell^6 a(T_o)^{4}} \Big) = 1 \ ,
\end{equation}
at the bounce point, respectively. In the latter case, the equation for $a(T_o)$ has to be solved numerically. For that, as the energy of electromagnetic field mode is small with respect to that of the clock field, one usually employs Newton method of root finding, setting its starting point as
\begin{equation}
    a_o(T_o) = \left( \frac{2\alpha_o^2 p_T}{3\pi G\ell^3} \right)^{\frac{1}{3}} \ .
\end{equation}

\bibliographystyle{apsrev4-1}
\bibliography{References}

\end{document}